# Nonstationary Stochastic Simulation of Strong Ground-Motion Time Histories: Application to the Japanese Database

**A. Laurendeau & F. Cotton**
*ISTerre, CNRS, Université Joseph Fourier, Grenoble, France*

**L-F. Bonilla**
*IFSTTAR, Paris, France*

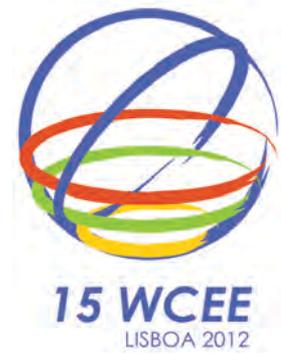

**SUMMARY:**
For earthquake-resistant design, engineering seismologists employ time-history analysis for nonlinear simulations. The nonstationary stochastic method previously developed by Pousse et al. (2006) has been updated. This method has the advantage of being both simple, fast and taking into account the basic concepts of seismology (Brune's source, realistic time envelope function, nonstationarity and ground-motion variability). Time-domain simulations are derived from the signal spectrogram and depend on few ground-motion parameters: Arias intensity, significant relative duration and central frequency. These indicators are obtained from empirical attenuation equations that relate them to the magnitude of the event, the source–receiver distance, and the site conditions. We improve the nonstationary stochastic method by using new functional forms (new surface rock dataset, analysis of both intra-event and inter-event residuals, consideration of the scaling relations and $V_{S30}$), by assessing the central frequency with S-transform and by better considering the stress drop variability.

## 1. INTRODUCTION

For earthquake-resistant design, engineering seismologists employ time-history analysis for nonlinear simulations. Within worldwide accelerometric databases, few natural records are available in the magnitude-distance range of engineering interest for low seismicity region (near-field and moderate magnitude scenarios). Consequently, it is necessary to develop simulation methods to generate synthetic accelerograms. Among the available techniques for generating time histories (Douglas and Aochi, 2008), we focus on methods based on stochastic simulations. These techniques have the advantage of using few input parameters (they do not require detailed knowledge about the rupture) and not being site specific but rather global. The recent development of a large strong-motion database is also a key motivation for the development of this type of method as the data availability helps to calibrate the input parameters (e.g. Pousse et al. 2006; Rezaeian and Der Kiureghian, 2010). We chose to update the nonstationary stochastic method initially developed by Sabetta and Pugliese (1996) and then by Pousse et al. (2006). This method has the advantage of being simple, fast and taking into account the basic concepts of seismology (Brune's source, realistic time envelope function, nonstationarity and ground-motion variability). The synthetic generation depends on only three input parameters: moment magnitude, source-site distance and site conditions; and on only three ground-motion parameters: Arias intensity (AI), significant relative duration ($D_{SR}$) and the evolution of the central frequency over time ($F_C(t)$). However, the functional forms used by Pousse et al. (2006) are too simple. In this view, a Japanese database including only surface rock recordings is built. Secondly, ground-motion prediction equations are derived from this rock-motion database for the 5 % damped acceleration response spectra (SA(f)), peak ground acceleration(PGA), AI, $D_{SR}$ and $F_C(t)$. Finally the semi-empirical nonstationary stochastic method is improved by considering the new functional forms and also by better taking into account the stress drop variability.

## 2. JAPANESE DATASET

After the destructive 1995 Kobe earthquake, Japanese scientists installed dense and uniform networks that cover the whole of Japan: the Hi-net (high-sensitivity), F-net (broadband), KiK-net and K-NET (strong-motion) networks (Okada et al. 2004). Each instrument is a three-component seismograph with a 24-bit analog-to-digital converter: the KiK-net and K-NET networks use 200-Hz and 100-Hz sampling frequencies. Japan is in a highly seismic area where a lot of quality digital data are recorded and made available to the scientific community (http://www.kyoshin.bosai.go.jp/indexen.html). In the present study, the KiK-net and K-NET strong-motion records are collected up to the end of 2009. To have consistent metaparameters, we only used events characterized in the F-net catalog. Thus, the values of $M_W$, the hypocenter location (latitude, longitude and depth), and the rake angle used for focal mechanism characterization, were determined by F-net. We fix 4.5 $M_W$ as the lower magnitude limit of our selection, and to select stations located on rock sites, we keep in the analysis only the stations with $V_{S30}$ greater than 500 m/s. In the case of the KiK-net network, with the sites characterized by velocity profiles ranging from 30 m to 2008 m, $V_{S30}$ can be computed. In the case of the K-NET network, the surveys are made down 20 m in depth. Using the KiK-net velocity models, Boore et al. (2011) provided equations that related $V_{S30}$ to $V_{SZ}$ for Z ranging from 5 m to 29 m in 1 m increments. $V_{S30}$ is estimated from these equations for the K-NET network. To include only crustal events, shallow events with a focal depth ≤ 25 km were selected. Offshore events were excluded but we chose to add the events with $M_W \geq 5.5$ in the sea west of the eastern border of Japan. A magnitude-distance filter was applied according to the Kanno et al. (2006) ground-motion prediction equation, which allows the data observed at large distances to be eliminated. We chose 2.5 gal as a PGA threshold. Following a visual inspection, faulty recordings like S-wave triggers, or recordings from multi-events are eliminated or shortened. At the same time, we collected all of the available fault-plane models for earthquakes with $M_W \geq 5.7$, as listed in Table 2.1. The source distance is the closest distance from a fault plane to the observation site, and it is the hypocentral distance in the case of earthquakes for which the fault model is not available. Our dataset finally consists of 2357 recordings, 405 observation sites (240 KiK-net and 165 K-NET) and 132 earthquake epicenters. The magnitude-distance distribution is shown in Fig. 2.1.

**Table 2.1.** Events for which the source geometry is taken into account to define the source-receiver distance ($R_{RUP}$). The Kagoshima 2 event has been described as two fault planes. In this case, the source-receiver distances are calculated for these two planes, and the shortest distance is selected.

| Name | Date | Mw | Strike | Dip | Length | Width | Reference |
|---|---|---|---|---|---|---|---|
| Kagoshima 1 | 199703261731 | 6.10 | 280 | 90 | 15 | 10 | Horikawa (2001) |
| Kagoshima 2 | 199705131438 | 6.00 | (280, 190) | (90, 90) | (9, 8) | (10, 10) | Horikawa (2001) |
| Yamaguchi | 199706251850 | 5.90 | 235 | 86 | 16 | 12 | Ide (1999) |
| Iwate | 199809031658 | 5.69 | 216 | 41 | 10 | 10 | Nakahara et al. (2002) |
| Tottori | 200010061330 | 6.62 | 145 | 90 | 28 | 17.6 | Ikeda et al. (2002) |
| Miyagi-Ken | 200307260713 | 6.10 | (220, 186) | (45, 52) | (6, 12) | (10, 10) | Miura et al. (2004) |
| Chuetsu | 200410231756 | 6.60 | 216 | 53 | 24 | 16 | Hikima and Koketsu (2005) |
| Chuetsu | 200410231803 | 5.90 | 20 | 34 | 8 | 8 | Hikima and Koketsu (2005) |
| Chuetsu | 200410231812 | 5.70 | 20 | 58 | 8 | 8.3 | Hikima and Koketsu (2005) |
| Chuetsu | 200410231834 | 6.30 | 216 | 55 | 20 | 12 | Hikima and Koketsu (2005) |
| Chuetsu | 200410271040 | 5.80 | 39 | 29 | 8 | 8 | Hikima and Koketsu (2005) |
| Rumoi | 200412141456 | 5.73 | 15 | 25 | 10 | 10 | Maeada and Sasatani (2009) |
| Fukuoka | 200503201053 | 6.60 | 123 | 87.7 | 32 | 28 | Kobayashi et al. (2006) |
| Noto-Hanto | 200703250942 | 6.70 | 58 | 66 | 30 | 18 | Momiyama et al., 2009 |
| Chuetsu-Oki | 200707161013 | 6.70 | 34 | 36 | 32 | 24 | Miyake et al. (2010) |
| Iwate-Miyagi | 200806140843 | 6.90 | 203 | 37 | 42 | 18 | Yokota et al. (2009) |

## 3. GROUND-MOTION PREDICTION EQUATIONS

Characterizing the ground-motion properties is an important issue for engineering seismology. In the present study, we derive models for the several key ground-motion parameters (PGA, SA(f), AI, $D_{SR}$, and $F_C(t)$).

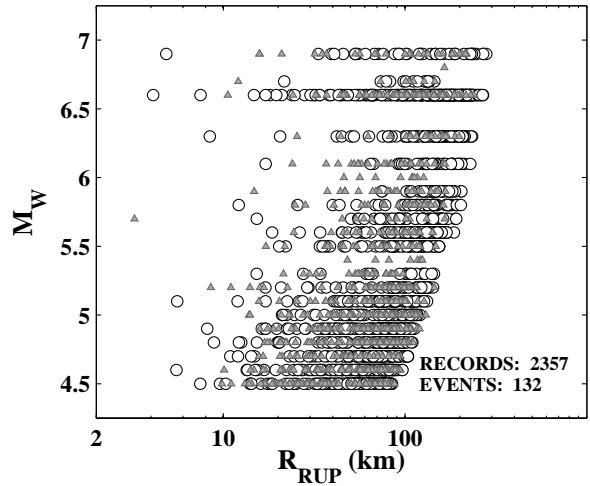

**Figure 2.1.** Distribution of moment magnitude ($M_W$) and rupture distance ($R_{RUP}$) of the selected records, differentiated by networks (o: KiK-net and Δ: K-NET networks).

## 3.1. Characterization of the ground-motion parameters

The Arias Intensity (AI) is a measure of the signal energy, as defined by Arias (1970). The definition of $D_{SR}$ is the time interval between 5% and 95% of the cumulative AI over time (Husid, 1969). The central frequency $F_C(t)$ quantifies the nonstationarity of the signal. It is related to the spectral moments of the Power Spectrum Density (Lai, 1982). The signal-to-noise ratio (SNR) is computed by using a 1-s window of noise. Therefore, only the records with at least 1 s of noise are used. $F_C(\tau)$ is computed for frequencies with SNR of more than 3, between the time arrival of the P wave and the time for which the AI reaches 95% by using S-transform (Stockwell et al., 1996). S-transform gives better resolution than the spectrogram used by Pousse et al. (2006), and especially for short-term recordings. The P waves are at higher frequencies than the S waves and most of the records tend to shift to lower frequencies as time increases. So statistically, $F_C(\tau)$ can be modelled as follows:

$$F_C(\tau) = \exp(A - B \times \ln(\tau + 1)) \qquad (3.1)$$

For each record, the A and B coefficients are assessed by a simple linear regression. However, the B coefficient is negative in some cases; this means that $F_C(\tau)$ increases globally. In this case, the A and B coefficients are computed both between the time of arrival of the P wave and the maximum of the S-transform, and between it and $D_{SR}(95\%)$. Finally, we take the values of the A and B coefficients from the S-wave part, or then the P-wave part if B is positive.

## 3.2. Functional forms

The ground-motion prediction equations are established for the two horizontal components. We consider different definitions (geometric mean and, arithmetic mean or independent components) for each parameter both to have a common definition for the ground-motion, and also to be consistent with previously published analyses. The ground-motion parameters are modelled as functions of the moment magnitude $M_W$, the closest distance from a fault plane to the observation site $R_{RUP}$ and a site parameter defined according to $V_{S30}$. In the present study, the aim was not to develop new functional forms. We then tested pre-existing functional forms, and we finally selected the functional forms that maximize the likelihood. The coefficients are derived from the random effects method (Abrahamson and Youngs, 1992), as well as from the within-event ($\phi$) and between-event ($\tau$) standard deviations of respectively $\delta W_{ij}$ and $\delta B_i$ residuals (the subscripts i and j refer to event and station).

*Peak ground acceleration and 5% damped acceleration response spectra:* Many studies have been conducted on these indicators. We chose to work with the following functional form (Boore and Atkinson, 2008; Rodriguez et al., 2011):

$$\ln(SA(f)_{ij}) = \begin{cases} a_1 + a_2 \cdot (M_{W_i} - M_h) + a_3 \cdot (M_{W_i} - M_h)^2 & \text{for } M_{W_i} \leq M_h \\ a_1 + a_4 \cdot (M_{W_i} - M_h) & \text{for } M_{W_i} \geq M_h \end{cases}$$
$$+ (b_1 + b_2 \cdot (M_{W_i} - 4.5)) \cdot \ln\left(\sqrt{R_{RUP_{ij}}^2 + h^2}\right) + b_3 \cdot \left(\sqrt{R_{RUP_{ij}}^2 + h^2}\right) \quad (3.2)$$
$$+ c_1 \cdot \ln(V_{S30_j}/800) + \delta W_{ij} + \delta B_{ij}$$

Some of the GMPE parameters are interdependent. Therefore some of them were constrained starting from subsets of data of Rodriguez et al. (2011). Deriving GMPE coefficients from a database with only $V_{S30} \geq 500$ m/s is not straightforward, as the number of records per event is more limited. So after trying to determine the coefficients $M_h$, $h$ and $b_3$ as reported by Rodriguez et al. (2011), we chose to use their coefficients because they allowed us to obtain a better distribution of residuals at different spectral periods.

*Arias Intensity*: Few functional forms have been developed for AI (Travasarou et al., 2003; Stafford et al., 2009; Foulser-Pigott and Stafford, 2012; Lee et al. 2012). Under the assumption of Parseval theorem, AI has been related to the acceleration Fourier amplitude spectrum of Boore (2003). Thus, Travasarou et al. (2003) and Stafford et al. (2009) deduct functional forms based on theory. On the other hand, various studies have shown strong correlations between PGA and AI (we found a coefficient of correlation of 0.90). In the way of Foulser-Pigott and Stafford (2012), we chose to adjust a functional form similar to PGA. Note that the functional forms developed by Travasarou et al. (2003) and Stafford et al. (2009) were also tested although they were finally selected since the form giving the best standard deviations is the one adapted from the PGA. This functional form is defined as:

$$\ln(AI_{ij}) = a_1 + a_2 \cdot (M_{W_i} - 5.6) + a_3 \cdot (M_{W_i} - 5.6)^2$$
$$+ (b_1 + b_2 \cdot (M_{W_i} - 4.5)) \cdot \ln\left(\sqrt{R_{RUP_{ij}}^2 + h^2}\right) \quad (3.3)$$
$$+ c_1 \cdot \ln(V_{S30_{ij}}/800) + \delta W_{ij} + \delta B_{ij}$$

*Significant relative duration*: Two recent papers have been published on this topic: Kempton and Stewart (2006) and Bommer et al. (2009). These studies developed duration functional forms from basic seismological theory (Boore, 2003), which predict that the source duration is inversely related to the corner frequency. We chose to adapt the form developed by Bommer et al. (2009):

$$\ln(D_{SR_{ij}}) = a_1 + a_2 \cdot (M_{W_i} - 5.6) + (b_1 + b_2 \cdot (M_{W_i} - 4.5)) \cdot \ln\left(\sqrt{R_{RUP_{ij}}^2 + h^2}\right)$$
$$+ c_1 \cdot \ln(V_{S30_{ij}}/800) + \delta W_{ij} + \delta B_{ij} \quad (3.4)$$

The $h$ coefficient controls the functional form at very short distances. However, the limited amount of rock data available in the near-field does not allow this coefficient to be calibrated. Therefore, we have chosen to use the $h$ coefficient determined previously by Bommer et al. (2009) on the NGA data.

*Central frequency over time:* Sabetta and Pugliese (1996), and then Pousse et al. (2006), developed simple functional forms for predicting the central frequency over time. Few studies have used this definition of central frequency for characterizing the frequency content in time. Indeed, engineering seismologists usually prefer to use the cumulative number of zero crossings over time (e.g. Rezaeian and Der Kiureghian, 2010). Pousse et al. (2006) showed that the A coefficient is mainly sensitive to site effects (A increases when $V_{S30}$ increases) and to the distance (A decreases when $R_{RUP}$ increases). This study also suggests that the B coefficient is mainly sensitive to the distance (the slope increases when $R_{RUP}$ decreases) and also to the magnitude (the slope increases when $M_W$ increases). Finally, we chose to develop a simple functional form as:

$$PAR_{ij} = a_1 + a_2 \cdot (M_{W_i} - 5.6) + b_1 \cdot \ln(R_{RUP_{ij}}) + c_1 \cdot \ln(V_{S30_{ij}}/800) + \delta W_{ij} + \delta B_{ij} \qquad (3.5)$$

with PAR= A, ln(B)

## 3.3. Results

The regression coefficients are presented in Tables 3.1 and 3.2. Figure 3.1 shows the inter-event and intra-event residuals according to $M_W$, $R_{RUP}$ and $V_{S30}$. For SA(f), AI and $D_{SR}$, the residuals are well distributed. For SA(f) and AI, the sigma deducted are stronger than those of previous studies. Our results show a standard deviation of 0.845, while Boore and Atkinson (2008) found a σ of 0.566 on NGA data and Rodriguez et al. (2011) found a σ of 0.78 on Japanese data. It was already recognized that the ergodic variability of ground-motion is stronger for Japan (Rodriguez et al., 2011), because of larger site variability. In the case of AI, our σ is also larger than those of previous studies (see Lee et al. (2012) for a summary of standard deviations achieved). For $D_{SR}$ prediction, our functional form is similar to that of Bommer et al. (2009) and the sigma obtained is similar to this previous study. However, in detail, the inter-event and intra-event terms show significant differences. Bommer et al. (2009) found the values of intra-event and inter-event almost similar, while our inter-event term is lower and our intra-event term is stronger (this confirms that Japanese sites might be more heterogeneous than the European and Californian sites for a given $V_{S30}$).

**Table 3.1.** Regression coefficients for PGA and 5 % damped acceleration response spectra (geometrical mean of the two horizontal components, g).

| Per. (s) | a1 | a2 | a3 | a4 | Mh | b1 | b2 | b3 | h | c1 | φ | τ | σ |
|---|---|---|---|---|---|---|---|---|---|---|---|---|---|
| PGA | -0.053447 | 0.51153 | -0.13258 | 0.22396 | 5.6 | -0.96551 | 0.2107 | -0.014 | 1.36 | -0.33707 | 0.65541 | 0.53346 | 0.84507 |
| 0.0384 | 0.66897 | 0.58703 | -0.20807 | 0.39223 | 5.6 | -1.0025 | 0.16005 | -0.014 | 1.2 | -0.062352 | 0.67452 | 0.5656 | 0.88027 |
| 0.0484 | 0.91612 | 0.64692 | -0.22739 | 0.49764 | 5.6 | -0.98952 | 0.13093 | -0.014 | 1.2 | -0.053157 | 0.69009 | 0.57479 | 0.89811 |
| 0.0582 | 1.0402 | 0.86293 | -0.15531 | 0.67188 | 5.6 | -0.931 | 0.08792 | -0.014 | 1.2 | -0.063994 | 0.70965 | 0.57829 | 0.91543 |
| 0.0769 | 1.174 | 1.1741 | -0.040642 | 0.82768 | 5.6 | -0.85675 | 0.049073 | -0.014 | 1.2 | -0.060246 | 0.75727 | 0.57423 | 0.95037 |
| 0.0844 | 1.1251 | 1.1062 | -0.077909 | 0.79292 | 5.6 | -0.85304 | 0.05869 | -0.014 | 1.2 | -0.15693 | 0.76705 | 0.56967 | 0.95545 |
| 0.097 | 1.049 | 1.1029 | -0.097397 | 0.83191 | 5.6 | -0.84279 | 0.060533 | -0.014 | 1.2 | -0.30829 | 0.76035 | 0.56389 | 0.94663 |
| 0.1167 | 0.95244 | 1.1226 | -0.033774 | 0.72045 | 5.6 | -0.86795 | 0.090565 | -0.0138 | 1.2 | -0.45209 | 0.73082 | 0.54812 | 0.91353 |
| 0.1472 | 0.9456 | 0.98923 | -0.056821 | 0.55631 | 5.6 | -0.957 | 0.1353 | -0.0131 | 1.2 | -0.63621 | 0.73475 | 0.53237 | 0.90735 |
| 0.1691 | 0.83211 | 0.6902 | -0.21985 | 0.3962 | 5.6 | -1.0205 | 0.18065 | -0.0126 | 1.2 | -0.70706 | 0.72299 | 0.52852 | 0.89557 |
| 0.2036 | 0.64394 | 0.4867 | -0.29401 | 0.25966 | 5.6 | -1.0727 | 0.21946 | -0.0119 | 1.2 | -0.76095 | 0.71194 | 0.52347 | 0.88367 |
| 0.234 | 0.47552 | 0.55449 | -0.30051 | 0.33502 | 5.6 | -1.0613 | 0.20849 | -0.0113 | 1.2 | -0.79769 | 0.70218 | 0.50345 | 0.86402 |
| 0.309 | 0.21697 | 0.40545 | -0.32381 | 0.21518 | 5.6 | -1.1407 | 0.25011 | -0.01 | 1.2 | -0.78078 | 0.67721 | 0.51352 | 0.84989 |
| 0.3551 | 0.26779 | 0.54112 | -0.12721 | -0.05996 | 5.8 | -1.1953 | 0.27496 | -0.0092 | 1.2 | -0.8036 | 0.67557 | 0.51987 | 0.85244 |
| 0.3896 | 0.23289 | 0.37994 | -0.17941 | -0.17986 | 6 | -1.2195 | 0.28412 | -0.0087 | 1.2 | -0.79823 | 0.67672 | 0.52827 | 0.8585 |
| 0.4274 | 0.15474 | 0.39176 | -0.15983 | -0.23609 | 6 | -1.2479 | 0.29566 | -0.0082 | 1.2 | -0.78573 | 0.67768 | 0.52833 | 0.85929 |
| 0.469 | 0.16342 | 0.50356 | -0.1262 | -0.19146 | 6 | -1.27 | 0.289 | -0.0076 | 1.2 | -0.75421 | 0.67615 | 0.53119 | 0.85985 |
| 0.5913 | 0.004647 | 0.53229 | -0.11926 | -0.20505 | 6 | -1.3329 | 0.3007 | -0.0062 | 1.2 | -0.69754 | 0.67022 | 0.54423 | 0.86336 |
| 0.7456 | -0.15406 | 0.72525 | -0.03463 | -0.21427 | 6 | -1.3808 | 0.30522 | -0.0049 | 1.2 | -0.69847 | 0.65969 | 0.51299 | 0.83568 |
| 0.818 | -0.16584 | 0.75843 | -0.025508 | -0.23741 | 6 | -1.4197 | 0.31173 | -0.0043 | 1.2 | -0.6784 | 0.65957 | 0.50082 | 0.82816 |
| 0.9401 | -0.28915 | 0.79785 | -0.042921 | -0.20283 | 6 | -1.4363 | 0.30944 | -0.0036 | 1.2 | -0.65954 | 0.66032 | 0.47773 | 0.81501 |
| 1.3622 | -0.80171 | 0.71254 | -0.048238 | -0.13681 | 6 | -1.5357 | 0.34884 | -0.002 | 1.2 | -0.66515 | 0.64908 | 0.41846 | 0.77228 |

**Table 3.2.** Regression coefficients for the ground-motion parameters. Different horizontal definitions have been used: GM, geometrical mean; AM, arithmetic mean and IND, the two horizontal components are independent.

| Per. (s) | H Def. | a1 | a2 | a3 | b1 | b2 | h | c1 | φ | τ | σ |
|---|---|---|---|---|---|---|---|---|---|---|---|
| AI (m/s) | AM | 7.90495 | 3.8684 | -0.15884 | -3.04157 | -0.24657 | 15.815 | -0.71102 | 1.17046 | 0.98146 | 1.5275 |
| AI (m/s) | GM | 7.92892 | 3.88485 | -0.15950 | -3.04614 | -0.24972 | 16.131 | -0.71189 | 1.16603 | 0.98209 | 1.5245 |
| SMD (s) | IND | 0.36220 | 0.34394 | X | 0.63582 | -0.038941 | 2.5 | -0.10385 | 0.43360 | 0.19766 | 0.4765 |
| SMD (s) | GM | 0.37827 | 0.33056 | X | 0.62982 | -0.036646 | 2.5 | -0.10080 | 0.42182 | 0.17488 | 0.4566 |
| A | GM | 3.55833 | -0.043563 | X | -0.17115 | X | X | 0.13792 | 0.33288 | 0.088269 | 0.34439 |
| B | GM | -1.01196 | 0.14835 | X | -0.24392 | X | X | -0.40941 | 0.97950 | 0.27920 | 1.01852 |

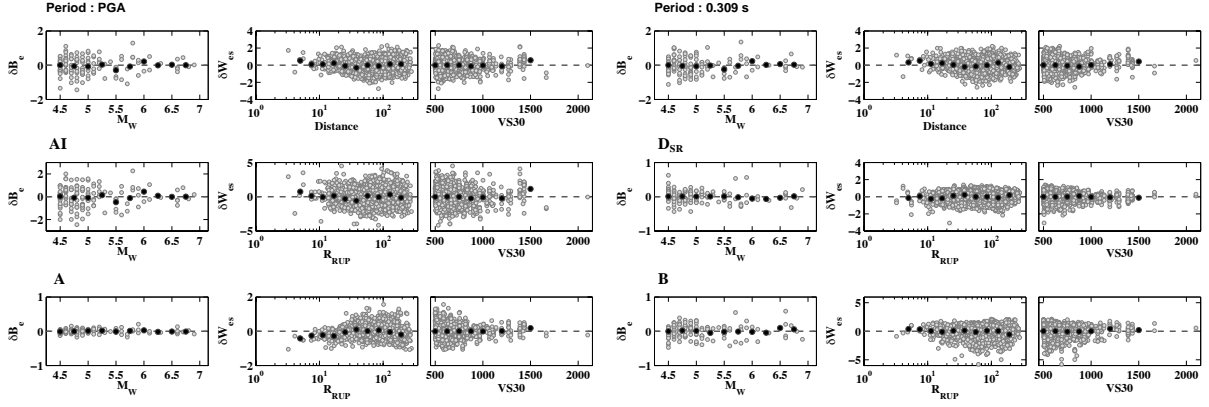

**Figure 3.1.** Inter-event residuals plotted with respect to $M_W$ and inter-event residuals as a function of $R_{RUP}$ and $V_{S30}$ at various spectral periods. The dark dots represent the average for a bin of $M_W$, $R_{RUP}$ or $V_{S30}$.

## 4. GENERATION OF TIME-HISTORIES WITH A NONSTATIONARY STOCHASTIC METHOD.

### 4.1. The nonstationary stochastic model formulation

Time domain simulations are derived from signal spectrogram $PS(t,f)$ which depends on the prediction of ground-motion parameters such as AI, $D_{SR}$ and $F_C(t)$ of the signal. Sabetta and Pugliese (2006) assumed that $PS(t,f)$ can be factorized as:

$$PS(t,f) = PS_t(f) \cdot Pa(t) \tag{4.1}$$

where $PS_t(f)$ represents the frequency content at each time t and $Pa(t)$ the time envelope function. This factorization needs that $PS_t(f)$ and $Pa(t)$ to follow a functional form based on the log-normal density of probability shape.

*Frequency content:* A benefit of this method is that its frequency content is realistic. It follows the ω-square model (Frankel et al., 1996), which is multiplied by a high-cut filter. This filter can account for the diminution of the high frequency motions (Boore, 2003). The nonstationarity is given by replacing $f_{max}$ with $F_C(t)$ in this filter. The seismic spectrum relative to the source can be defined as:

$$S_\tau(f) = \frac{\dfrac{(2\pi f)^2}{1+(f/f_c)^2}}{\sqrt{1+\left[f/F_{C_\tau}\right]^8}} \tag{4.2}$$

where $f_c$ is the corner frequency defined as: $\log(f_c) = 1.341 + \log(\beta \times \Delta\sigma^{1/3}) - 0.5 M_W$ where $\Delta\sigma$ is the stress drop in bars, $M_W$ is the moment magnitude and $\beta$ is the shear wave velocity in kilometers per second (see Table 4.1). $PS_t(f)$ is assumed to follow the functional form of a probability density function, so its area must be 1 and its dimension should be that of a duration (s). $S_\tau(f)$ must be normalized by its area. Moreover, for each simulation, the stress drop and the central frequency are drawn randomly. The Pousse et al. (2006) method did not correctly take into account this stress drop dependency, and we have improved this part. The simulations are now calibrated to a stress drop reference of 10 bars (Fig. 4.1).

*Envelope time function*: In the literature, different temporal envelopes are used (a gamma distribution a lognormal distribution). Only the time envelope of Pousse et al. (2006) includes the arrival of P, S, and coda waves; and the radiated energy is distributed over each specified duration. The only difference between our method and the former Pousse et al. (2006) studies concerns the evaluation of the quality factor, $Q=Q_0 f^N$ (see Table 4.1). In Pousse et al. (2006), the frequency was chosen randomly between 1 Hz and 5 Hz, whatever the chosen scenario. Now, the frequency is the central frequency predicted for a given scenario.

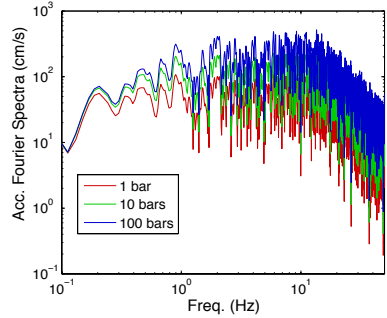

**Figure 4.1.** Simulated acceleration Fourier spectra for different values of stress drop.

*Variability:* Another key advantage of this simulation method is to take into account the natural variability of ground motion. A Monte Carlo exploration of the key parameters is used to reproduce this variability (see Tab. 4.1).

**Table 4.1.** Parameters used to simulate variability with the nonstationary stochastic method.

| Parameter | Fitted distribution | Distribution Bounds | Reference |
|---|---|---|---|
| Phase ($\varphi$) | Uniform | $[-\pi ; \pi]$ | |
| $\Delta\sigma$ (bars) | Uniform | $\text{Log}([0;2])$ | Lay and Wallace (1995) |
| $\beta$ (km/s) | Fixed | 3.6 | |
| $Q_0$ | Uniform | [ 45 ; 140 ] | Oth et al. (2011) |
| N | Uniform | [ 0.5 ; 0.9 ] | |
| $\ln(AI)$ | Normal | $\mu$ and $\sigma$ results from GMPEs | |
| $\ln(D_{SR})$ | | | |
| A and $\ln(B)$ | | | |

*Computation of the time history:* The accelerogram calculation is based on the assumption that the ground-motion at time t results from the contribution of random and uncorrelated phases (Boore, 2003). Synthetic accelerograms are then performed, by summing Fourier series with time-dependent coefficients derived from PS(t,f), as follows:

$$x(t) = \sum_{n=1}^{N} C_n(t) \cos(n 2\pi df t + \varphi_n) \tag{4.6}$$

$$C_n(t) = \sqrt{2PS(t,f)df} \tag{4.7}$$

where x(t) is the simulated accelerogram and the phases $\varphi$ are random numbers uniformly distributed in the range $-\pi$ to $\pi$.

## 4.2. Results

We compare the simulations obtained with real data. The chosen scenarios are all sampled by a significant number of records, i.e. a first scenario with $M_W = 5$, $R_{RUP} = 50$ km and $V_{S30} = 550$ m/s, and a second scenario with $M_W = 6.6$, $R_{RUP} = 30$ km and $V_{S30} = 550$ m/s. This method has the advantage that it can simulate many accelerograms in a few minutes. 2500 time histories were generated for the two scenarios by exploring the strong motion parameters between $\pm 1$ $\sigma$. Figure 4.2 shows a comparison between the predicted response spectrum and the simulated response spectra. Until about 0.3 s, the distribution of the spectral amplifications is well reproduced by the simulations. However, at long periods, the simulated response spectra tend to overestimate the spectral amplitudes. Within these simulations, a subset has been selected. The mean squared error (MSE) between the spectral accelerations of the simulations and the target spectrum (the mean response spectrum in this example), were computed. Figure 4.2 also shows the 30 time histories that have the best match with the target spectrum over the whole period range. Note that using this method, the selected time histories show a better fit at long periods. Figure 4.3 compares the distributions of the ground-motion parameters from the predicted model with the simulations. For these three indicators, the variability is well reproduced by the simulations. Figure 4.4 shows for comparison a sample of simulated and real time histories. The simulated time histories show different waveforms, various amplitudes, and durations like the real time histories.

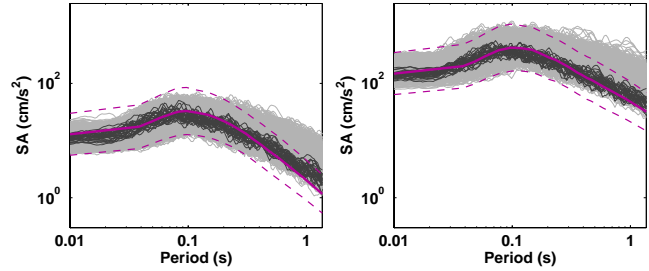

**Figure 4.2.** Comparison of simulated (gray lines) and predicted (purple line) acceleration response spectra corresponding to (left) $M_W$=5, $R_{RUP}$=50km and $V_{S30}$=550m/s; and (right) $M_W$=6.6, $R_{RUP}$=30km and $V_{S30}$=550m/s. In light gray lines, the 2500 simulations and in dark gray lines, the 30 best matched time-histories.

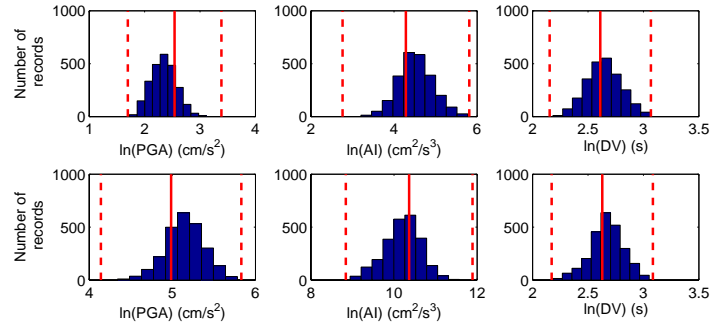

**Figure 4.3.** Comparison between the predicted distributions of the ground-motion parameters (red lined) and the ground-motion parameters distributions of the 2500 simulations (histogram) for the two scenarios: (top) $M_W$=5, $R_{RUP}$=50km and $V_{S30}$=550m/s; and (bottom) $M_W$=6.6, $R_{RUP}$=30km and $V_{S30}$=550m/s.

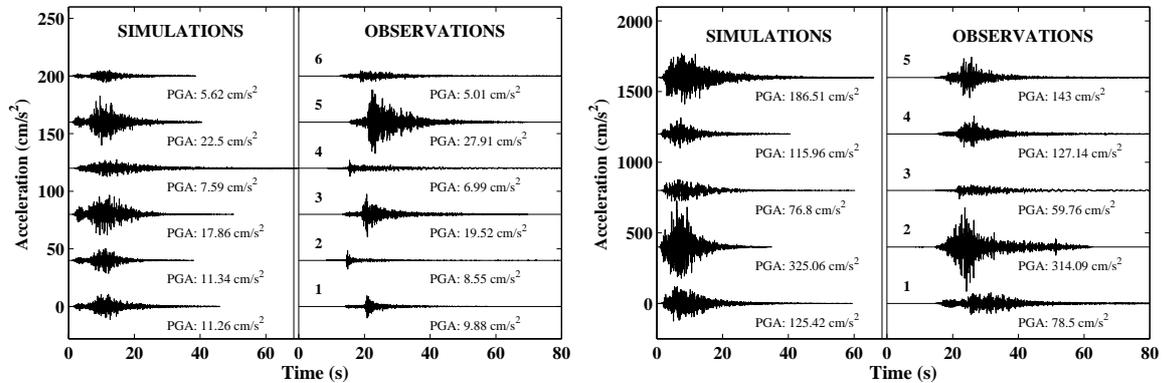

**Figure 4.4.** Comparison of simulated and recorded acceleration motions: (left) $M_W$=5, $R_{RUP}$=50 ± 1 km and $V_{S30}$=550 ± 50 m/s; and (right) $M_W$=6.6, $R_{RUP}$=30 ± 5 km and $V_{S30}$=550 ± 50 m/s.

## 5. CONCLUSIONS AND DISCUSSION

A new and homogeneous active shallow crustal accelerometric database of rock ground motion was built ($V_{S30} \leq 500$ m/s). This database consists of 4998 Japanese digital records.

New, rock-specific, ground-motion prediction equations were derived using this new database for the characterization of the key parameters (PGA, SA(f), AI, $D_{SR}$ and $F_C(t)$). These new GMPEs take into account recent developments (analysis of both intra-variability and inter-variability, updated functional forms, including the scaling relations and $V_{S30}$). Our equations should be used only for predictor variables in these ranges: $4.5 \leq M_W \leq 6.9$, and $500 \leq V_{S30} \leq 1500$ m/s. The intra-event is stronger than that derived in other regions, confirming that Japanese sites might be more heterogeneous than European and Californian sites. In addition, the site coefficient of these new rock specific GMPEs (PGA, AI and $D_{SR}$) was found to be smaller than that predicted from previous studies mixing rock and soil records. This shows the importance of having developed a specific rock database.

The nonstationary stochastic method is a simple method that includes the theoretical bases of seismology. This method allows the rapid generation of a large number of time histories. The new functional forms of the key ground-motion parameters and a new method to take into account the stress drop variability have been included. This method allows good reproducing of the ground motion and its variability at low periods, up to around 0.3 s. However, the response spectra are overestimated at long periods and so, the displacement is overestimated. This overestimation might have several origins:

- As reported by Safak and Boore (1988) and Rezaeian and Der Kiureghian (2010), it is necessary to assure zero residual velocity and displacement of the motion, as well as realistic response spectral values at long periods. Without such filtering, stochastically generated ground motions tend to overestimate response spectral values in the long period range. A high-pass filter was applied by these authors. However, our nonstationary stochastic method still includes a filter above the corner frequency (the ω-square model) and the use of this high-pass filter leads to an underestimation of the low frequencies.

 -As reported by Atkinson and Silva (2000), the Brune point-source model leads to an overestimation of the long periods for large magnitude events. They have shown that the use of a stochastic finite-fault model or the use of a two-corner point-source model for the earthquake spectrum allow this overestimation to be eliminated. These two corner frequencies are however empirically derived, and they are difficult to apply outside the specific context of California.

## ACKNOWLEDGEMENTS
This work was funded by the CASHIMA project (ILL and CEA). We thank Fabrice Hollender (CEA) and Vérinique Caillot (ILL) for their support and useful suggestions. We are indebted to the National Research Institute for Earth Science and Disaster Prevention (NIED), Japan, for providing the data for this analysis and Hadi Ghasemi for providing the fault plane models. We have benefited from the experience of Adrian Rodriguez, Gonzalo Montalva, Sinan Akkar and Abdullah Sandikkaya regarding the regression procedure.